\documentclass[journal=langd5,manuscript=article]{achemso}
\DeclareSymbolFont{matha}{OML}{txmi}{m}{it}% txfonts
\DeclareMathSymbol{\varv}{\mathord}{matha}{118}
\usepackage[version=3]{mhchem} % Formula subscripts using \ce{}
\usepackage{amssymb}
\usepackage{xcolor}
\author{Panagiotis E. Theodorakis}
\affiliation{Institute of Physics, Polish Academy of Sciences, Al. Lotnik\'ow 32/46, 02-668 Warsaw, Poland}
\email{panos@ifpan.edu.pl}
\author{Alidad Amirfazli}
\affiliation{Department of Mechanical Engineering, York University, Toronto, ON, M3J 1P3, Canada}
\author{Bin Hu}
\affiliation{Flow Capture AS, Industriveien 1, 2020 Skedsmokorset, Norway}
\author{Zhizhao Che}
\affiliation{State Key Laboratory of Engines, Tianjin University, 300072 Tianjin, China}

   \title[Droplet Control Based on Pinning and Substrate Wettability]
  {Droplet Control Based on Pinning and Substrate Wettability}


\begin{document}

\begin{tocentry}

\includegraphics[width=8.25cm]{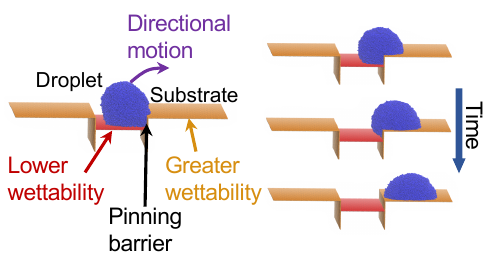}

\end{tocentry}

\begin{abstract}
Pinning of liquid droplets on solid substrates is ubiquitous and plays an essential role in many
applications, especially in various areas, such as microfluidics and biology. Although pinning
can often reduce the efficiency of various applications, a deeper understanding of this 
phenomenon can actually offer possibilities for technological exploitation. Here, by means of
molecular dynamics simulation, we identify the conditions that lead to droplet pinning or
depinning and discuss the effects of key parameters in detail, such as the height of the physical
pinning-barrier and the wettability of the substrates. Moreover, we describe the mechanism of the
barrier crossing by the droplet upon depinning, identify the driving force of this process, and,
also, elucidate the dynamics of the droplet. Not only does our work provide a detailed
description of the pinning and depinning processes, but it also explicitly highlights how both
processes can be exploited in nanotechnology applications to control droplet motion. Hence, we
anticipate that our study will have significant implications for the nanoscale design of
substrates in micro and nano-scale systems and will assist with assessing pinning effects in
various applications.
\end{abstract}

\vspace{0.7in}

\section{INTRODUCTION}
The control of droplets on solid substrates is crucial for many applications in various areas,
such as microfluidics, microfabrication, coatings, and biology. To this end, the accurate
steering of droplets' motion can be realised by proper substrate design. In materials science, 
for example, a design based on micro-pillar structures has been shown to lead to superhydrophobic
substrates \cite{Lafuma2003} for, among others, self-cleaning \cite{Min2008} and anti-icing
\cite{Antonini2011}. As a result of this specific design, pinning effects naturally arise that
may affect droplet's motion by introducing a sticky or slippery behaviour
\cite{Huang2019,Wu2014,Li2018,Zheng2007,Chen2016}, which also depends on substrate wettability
\cite{Antonini2012,Chen2015}. By means of lubrication theory, Joanny and Robbins have
investigated the dynamics of a contact line on a heterogeneous plate, which is advanced at
constant force or velocity \cite{Joanny1990}. They have unveiled the scaling of the force and the
velocity and, also, found that alternating patches of constant wettability produce a linear
relation. 
Esp\'in and Kumar have presented a model based on lubrication-theory to describe
contact-line pinning on substrates with heterogeneities. The work has discussed the effect 
of roughness  through a continuum model that has shown to agree with experiments.\cite{Espin2015}  
Alava and Dub\'e have analysed the statistical properties of the spreading contact line 
(droplet radius and contact angle) on heterogeneous surfaces.\cite{Alava2012}
Moreover, Marmur has described the equilibrium wetting on rough surfaces determining the transition
between homogeneous and heterogeneous wetting regimes on the basis of the Wenzel and Cassie--Baxter
equations.\cite{Marmur2003}
Experimentally, Ramos and Tanguy have studied the pinning--depinning
phenomenon of a contact line on a solid surface decorated by a random array of nanometric
structures and found a linear relation between the hysteresis caused by defects and their areal
density \cite{Ramos2006}. In this context, the relation between the dynamic contact angle and
contact line speed has been recently considered by numerical simulation \cite{Yamamoto2016}. 
In another example,
substrates characterised by a gradient of a physical or a chemical property in a particular
direction along the substrate can steer the motion of liquid droplets without the requirement of
an external energy source \cite{Zhang2019a,Li2018a,Leng2020,Bardall2020, Zhao2020}. A well-known
example is durotaxis, where a droplet can autonomously move along a substrate due to the presence
of a stiffness gradient \cite{Barnard2015,Chang2015,Theodorakis2017, Moriyama2018}, which 
crucially depends on the wettability of the substrate \cite{Theodorakis2017}. In any of the above
systems, pinning of contact line can be advantageous or impede droplet motion or its
manipulation, leading to a greater or lower efficiency of relevant processes
\cite{Zhang2019, Huet2020,Al-Sharafi2018,Huang2019,Wu2014,Orejon2011,Ondarcuhu1995,Pham2017}. 
There are still outstanding issues that remain regarding the possibility of exploiting the effects
of droplet pinning and substrate wettability in controlling droplet's motion. This is especially 
true regarding microlevel origins of pinning and its mechanism, which can be advantageous for
various nanotechnology applications.

This paper aims at filling the above gap by taking advantage of high-fidelity \textit{in silico}
experiments at nanoscale. We employ molecular dynamics (MD) simulation based on a coarse-grained
model and the system setup of Figure~\ref{fig:1}. Apart from aiming at acquiring an in-depth
understanding of droplet pinning on solid substrates with different wettability, we also argue
that the pinning has the potential of controlling nanodroplets, for example, selective droplet
separation. For this reason, we have studied a range of different pinning scenarios, which
include various combinations of substrate wettabilities and pinning barriers for droplets of
different sizes. Thus, we anticipate that our results will inspire the design of substrates for
steering droplets in micro- and nano-scale systems and will assist with assessing pinning
effects in a range of different nanotechnological applications.

\section{MATERIALS AND METHODS}

\begin{figure}[bt!]
\centering
 \includegraphics[width=0.5\linewidth]{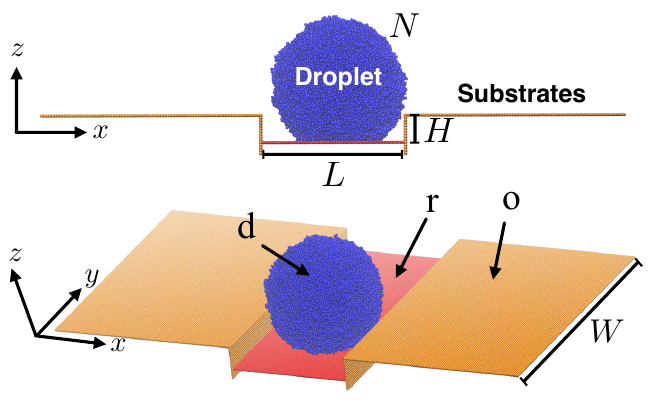}
\caption{\label{fig:1} A typical initial configuration of our simulations. Two different views
of the same configuration are presented in the upper and the lower panels for the sake of clarity.
The system consists of substrates with different wettability indicated by different colours
(red (r) and orange (o)). The droplet (d) consists of blue beads and is placed onto the red
substrate. The pinning barrier is characterised by the height, $H$, between the parallel (along the
$x-y$ plane) red and orange substrates. Different values of $H$ and different wettabilities for the red and
orange substrates are considered in this study. The length, $L$, depends on the size of the droplet
and the wettability of the red substrate, and is large enough to guarantee that the droplet is in a
state as the one illustrated in the upper panel with an appropriate distance between the droplet and
the lateral orange substrate (perpendicular to the $x-y$ plane). The width, $W$ is chosen to guarantee
that mirror images of the droplet in the $y$ direction are not interacting, due to the
periodic boundary conditions, which are applied in all directions. Here, the example
refers to a droplet with $N=50400$ coarse-grained beads, $H=12\sigma$, and interaction between the
droplet and the red substrate $\varepsilon_{\rm rd}=0.3\varepsilon$. Snapshots have been produced
with the VMD software \cite{Humphrey1996}. }
\end{figure}

We have used MD simulations of a coarse-grained model \cite{Theodorakis2017,Kremer1990}
where interactions between different components of the system, \textit{i.e.} the drop
and the substrate beads, are described by means of the Lennard-Jones (LJ) potential, namely,

\begin{equation}\label{eq:LJpotential}
U_{\rm LJ}(r) = 4\varepsilon_{\rm ij} \left[  \left(\frac{\sigma_{\rm  ij}}{r}
\right)^{12} - \left(\frac{\sigma_{\rm ij}}{r}  \right)^{6}    \right],
\end{equation}
where $r$ is the distance between any pair of beads in the system, and $i$ and $j$ indicate the type
of beads: `d' for droplet beads, `r' for the beads that belong to the red substrate, and `o' for the
beads of the orange substrates (Figure~\ref{fig:1}). In our model, $\sigma_{\rm ij} = \sigma$ for all
combinations of types $\rm i$ and $\rm j$, with $\sigma$ being the unit of length. As usual, the LJ
potential is cut and shifted at a cutoff distance $r_{c}=2.5\sigma$ for any interaction involving 
the droplet beads, while $r_{c}=2^{1/6}\sigma$ (purely repulsive potential) for any interactions
between the substrate beads. The strength of the interactions is defined by the parameter
$\varepsilon_{\rm ij}$ of the LJ potential. In our case, the parameters, $\varepsilon_{\rm rd}$ and
$\varepsilon_{\rm od}$ vary between 0.3$\varepsilon$ and 0.7$\varepsilon$, where $\varepsilon$ is 
the energy unit and $k_B$ (Boltzmann's constant) is considered as unity \cite{Theodorakis2017}.
The interactions $\varepsilon_{\rm rd}$ and $\varepsilon_{\rm od}$ are used to tune the
wettability of the droplet on the red and the orange substrates (Figure~\ref{fig:1}).

We have considered droplets of different size, which consist of $N=112$, $1008$, or $5040$ 
chains of ten coarse-grained beads each. The finite extensible nonlinear elastic (FENE) 
potential \cite{Kremer1990} was used to tether together consecutive beads in these polymer chains,
which is mathematically expressed as follows:

\begin{equation}\label{eq:KG}
 U_{\rm FENE}(r) = -0.5 K_{\rm FENE} R_{\rm 0}^2 \ln \left[ 1 - \left(\frac{r}{R_{\rm 0}} \right)^2  \right],
\end{equation}
where $r$ is the distance between two consecutive beads along the polymer backbone, 
$R_{\rm 0}=1.5\sigma$ expresses the maximum extension of the bond, and
$K_{\rm FENE} = 30 \varepsilon/\sigma^2$ is an elastic constant. For the chosen chain length, there
aren't any evaporation effects and the vapour pressure is therefore sufficiently 
low \cite{Tretyakov2014,Theodorakis2017}.

To evolve our system in time, we used MD simulation by choosing the Langevin thermostat
\cite{Schneider1978} as implemented in the LAMMPS package \cite{Plimpton1995}. The time unit in 
our simulations is $\tau =\sqrt{m\sigma^2/\varepsilon}$, where $m$ is the mass unit. The 
time-step for the integration of the equations of motion for the droplet particles is 
$\Delta t =0.005\tau$. Thus, the temperature $T$ fluctuates around a predefined value
$T=\varepsilon/k_{B}$, where $k_{B}$ is the Boltzmann constant, and the energy $\varepsilon$ is
measured in units of $k_B T$. Periodic boundary conditions are applied in all directions and we
guarantee that mirror images of the droplet do not interact with each other in any direction. 
A typical initial configuration for our systems is illustrated in Figure~\ref{fig:1}. Typical
trajectories for our systems start from such initial configurations. We have run simulations up
to $10^8$ MD time steps for cases that remained pinned to ensure that unpinning will not happen 
at a very late time of the simulation. For droplets that cross the pinning boundary, the length 
of the trajectories was up to the point that the droplet reached the final equilibrium state on
top of the orange substrate. Our results are based on the analysis of these trajectories.

\section{RESULTS AND DISCUSSION}

Before delving into the details of the system, it should be mentioned that pinning can be the result
of chemical inhomogeneity, surface roughness (or a physical step), or a combination of both. In this
work, we will consider the combined effect of physical barrier and wettability to allow for a 
comprehensive understanding.
Pinning is defined as inability of the contact line to move; such inability is rooted in
the thermodynamic energy barrier due to chemical and/or physical heterogeneity expressed on a surface. 
In this study, such barrier to movement of the contact line is through the physical barrier that 
prevents the droplet from moving on top of the orange substrate; the wettability of the physical
heterogeneity is also varied. Due to the attractive nature of the LJ interaction, the droplets 
in this  study are pinned at the boundary between the red and orange substrates, as such the 
pinning inherently takes place without imposing a pinning requirement.
The system studied here consists of a droplet on a substrate that is
parallel to the $x-y$ plane, as shown in Figure~\ref{fig:1}. The wettability of the substrate by
the droplet is determined by the Lennard-Jones (LJ) interaction-parameter, $\varepsilon_{\rm rd}$,
where `r' indicates the red colour of the substrate and `d' the droplet (Figure~\ref{fig:1}). 
A larger value of $\varepsilon_{\rm rd}$ allows
for a higher wettability of the substrate, whereas a smaller value corresponds to a lower
wettability. From our previous study \cite{Theodorakis2017}, the choice,
$0.3\varepsilon \leq \varepsilon_{\rm rd} \leq 0.7\varepsilon$, maintains the spherical-cap 
shape of the droplet on a substrate monolayer and avoids evaporation effects and
large distortions of the droplet contact line.
In this case, the contact angle of the droplet is uniquely defined by the 
strength of the LJ interaction (\textit{e.g.} $\varepsilon_{\rm rd}$) and linearly depends on it.
\cite{Theodorakis2015} In particular, LJ energy parameters in the range $0.3 - 0.7\varepsilon$ would yield
contact angles in the range 60$^\circ$--120$^\circ$.\cite{Theodorakis2017}
In addition, two orange substrates 
perpendicular to the $x-y$ plane and two orange
substrates parallel to the $x-y$ plane are part of the same system as illustrated in
Figure~\ref{fig:1}. Both orange substrates have the same wettability, which is expressed by the
interaction strength of the LJ potential, $\varepsilon_{\rm od}$, where `o' stands for the orange
colour of the substrates. The orange substrates, which are parallel to the $x-y$ plane, and the red
substrate are separated by a distance, $H$, in the $z$ direction, which corresponds to the height
of the physical barrier that the droplet needs to overcome in order to move from the red substrate
to the orange substrate. The pinning barrier, namely the height, $H$, can
vary by changing the position of the red substrate in the $z$ direction. 
The choice of lengths, $L$ and $W$ (Figure~\ref{fig:1}), does not affect our results.
$L$ is chosen such that the droplet sticks to the pinning barrier after a short time, since
the interaction of the droplet with the red and the orange substrates is always attractive. 
 $W$ is large enough to guarantee that mirror images of the 
droplet do not interact in the $y$ direction due to the presence of the periodic boundary conditions.
Hence, depending on the
choice of the parameters, $H$, $\varepsilon_{\rm rd}$, and $\varepsilon_{\rm od}$, as well as the
droplet size (total number of beads, $N$), the droplet may be able to overcome 
(cross) the pinning barrier
and potentially reach a new equilibrium state on top of one of the orange
substrates. In the following, we discuss the effects of these parameters on droplet pinning and
describe the mechanism of droplet motion over the barrier upon droplet depinning.

\begin{figure*}[bt!]
 \includegraphics[width=\linewidth]{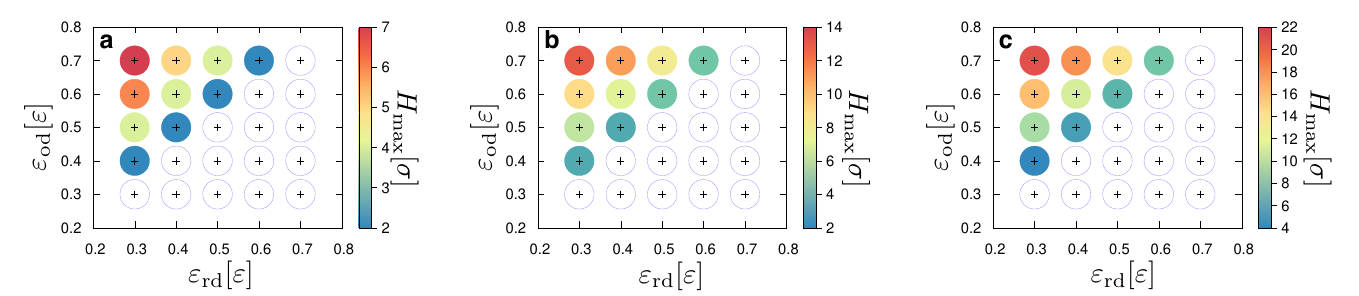}
\caption{\label{fig:2} Maximum height of the pinning barrier, $H_{\rm max}$, that a droplet is able to
overcome for different values of wettability for the red and the orange substrates as expressed via
the parameters $\varepsilon_{\rm rd}$ and $\varepsilon_{\rm od}$, respectively. The black crosses
indicate the exact values of $\varepsilon_{\rm rd}$ and $\varepsilon_{\rm od}$ on the graphs for
each circle. The colour code reflects the value of $H_{\rm max}$ for each set of parameters. The
empty circles indicate that the droplet remains pinned (on the red substrate) for any value of
$H>=1.0\sigma$. Each plot refers to droplets of different size, namely $N=1120$ (a), $N=10080$ (b),
and $N=50400$ (c) beads.}
\end{figure*}

Figure~\ref{fig:2} presents the results on the maximum height of the pinning barrier, $H_{\rm max}$, that
the droplet is able to overcome.
In particular, the dependence of $H_{\rm max}$ on the parameters $\varepsilon_{\rm rd}$ and
$\varepsilon_{\rm od}$ for droplets of different sizes is laid out. We observe that the
droplet will remain pinned, when the red substrate has a greater wettability than
the orange substrates, independently of the droplet size. In other words, $\varepsilon_{\rm od}$
must always be larger than $\varepsilon_{\rm rd}$ to allow for droplet depinning. Hence, the thermal
fluctuations of the droplet alone are not sufficient to enable depinning, even for values of $H$ as
low as $H=\sigma$, and even for our largest droplets ($N=50400$ beads). However, droplets can
generally overcome ever larger barriers as their size increases when $\varepsilon_{\rm od}>\varepsilon_{\rm rd}$
 and for the range of values considered in this study. In particular, $H_{\rm max}$ can be as high as
$21\sigma$ in the case of a droplet consisting of $N=50400$ beads (Figure~\ref{fig:2}c, 
$\varepsilon_{\rm rd}=0.3\varepsilon$ and $\varepsilon_{\rm od}=0.7\varepsilon$). In contrast, a droplet of $N=1120$
beads would only overcome a barrier of $7\sigma$ at best (Figure~\ref{fig:2}a, 
$\varepsilon_{\rm rd}=0.3\varepsilon$ and $\varepsilon_{\rm od}=0.7\varepsilon$). Moreover, the
value of $H_{\rm max}$ crucially depends on the  wettability difference between the red and
the orange substrates in each case, as expressed through the LJ parameters $\varepsilon_{\rm rd}$ and
$\varepsilon_{\rm od}$. In particular, the larger the difference in wettability, the larger the
$H_{\rm max}$ the droplet is able to overcome. In other words, as the difference in wettability between
the red and orange substrates becomes smaller, $H_{\rm max}$ decreases. In addition,
choosing the highest
possible wettability for the orange substrates always yields the largest $H_{\rm max}$, which suggests
that maximising $\varepsilon_{\rm od}$ favours droplet depinning. For example, the combination
($\varepsilon_{\rm rd}=0.5\varepsilon$, $\varepsilon_{\rm od}=0.7\varepsilon$) results in a larger value
of $H_{\rm max}$ in comparison with the combination ($\varepsilon_{\rm rd}=0.3\varepsilon$,
$\varepsilon_{\rm od}=0.5\varepsilon$) in the case of all droplet sizes, despite the absolute difference between the parameters
$\varepsilon_{\rm rd}$ and $\varepsilon_{\rm od}$ being the same. Eventually, the affinity of the 
droplet to the orange substrates drives the crossing of the barrier, as will be discussed further below. In summary, the largest $H_{\rm max}$ is
achieved for ($\varepsilon_{\rm rd}=0.3\varepsilon$, $\varepsilon_{\rm od}=0.7\varepsilon$) and the
smaller $H_{\rm max}$ for the combination ($\varepsilon_{\rm rd}=0.3\varepsilon$,
$\varepsilon_{\rm od}=0.4\varepsilon$). In view of these observations, we present in the following results of
pinning and depinning (see also movies in Supplementary Information) by keeping 
$\varepsilon_{\rm rd}=0.3\varepsilon$ constant, and varying
$\varepsilon_{\rm od}$, as well as results where we keep $\varepsilon_{\rm od}=0.7\varepsilon$
constant, and vary $\varepsilon_{\rm rd}$.

\begin{figure*}[bt!]
\centering
 \includegraphics[width=\textwidth]{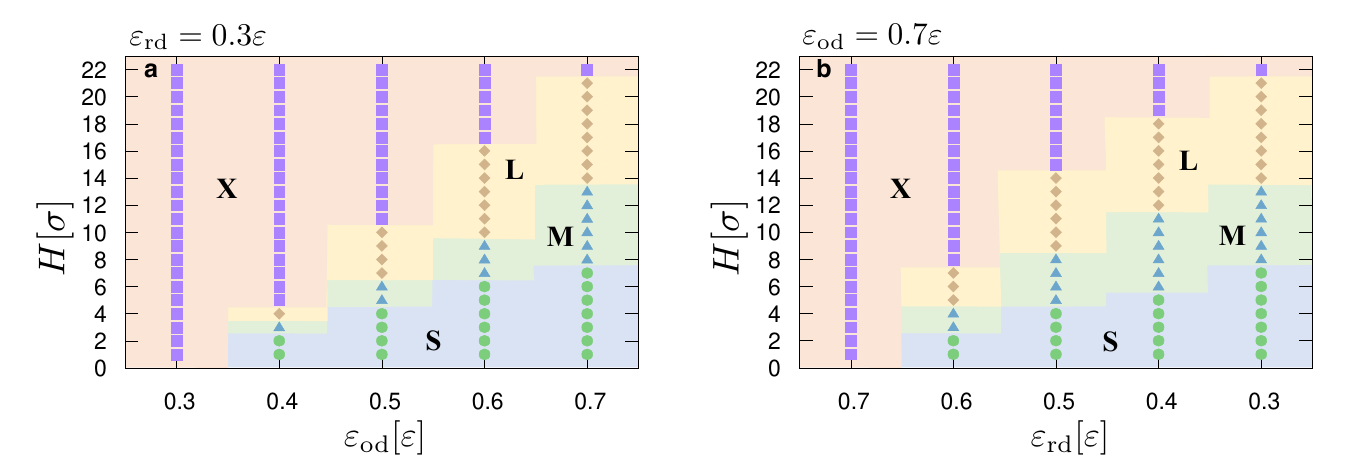}
\caption{\label{fig:3} (a) State diagram (pinning/depinning) for droplets of different size (small-size
droplets with $N=1120$ beads, medium-size droplets with $10080$ beads, and large-size droplets with
$50400$ beads) as a function of the height, $H$ ($y$ axis), and the interaction parameter
$\varepsilon_{\rm od}$ ($x$ axis). $\varepsilon_{\rm rd}=0.3\varepsilon$. `S' indicates cases (circles)
for which small, medium and large droplets are able to overcome a barrier, $H$. `M' indicates cases
(triangles) for which only the medium and the large droplets are able to overcome a pinning barrier of
height $H$. `L' indicates cases (diamonds) for which only the large droplets can overcome a pinning
barrier of height $H$. Finally, `X' indicates cases (squares) for which the droplets remain pinned,
irrespective of their size. (b) Similar to panel (a), but results refer to cases where
$\varepsilon_{\rm od}=0.7\varepsilon$ and $\varepsilon_{\rm rd}$ varies ($x$ axis), as indicated on
the graph.}
\end{figure*}

\begin{figure*}[bt!]
 \includegraphics[width=\linewidth]{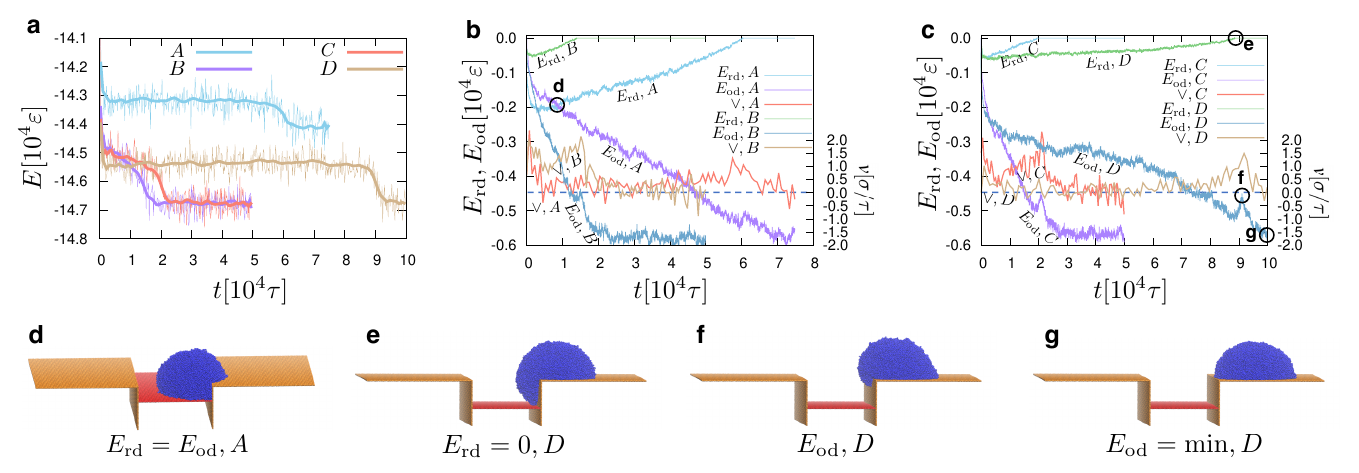}
\caption{\label{fig:4} (a) Total potential energy, $E$, of four different systems (A, B, C, and D) based
on a droplet with $N=50400$ beads and parameters,
A: $H=12\sigma$, $\varepsilon_{\rm rd}=0.5\varepsilon$, and $\varepsilon_{\rm od}=0.7\varepsilon$;
B: $H=12\sigma$, $\varepsilon_{\rm rd}=0.3\varepsilon$, and $\varepsilon_{\rm od}=0.7\varepsilon$;
C: $H=16\sigma$, $\varepsilon_{\rm rd}=0.3\varepsilon$, and $\varepsilon_{\rm od}=0.7\varepsilon$;
D: $H=21\sigma$, $\varepsilon_{\rm rd}=0.3\varepsilon$, and $\varepsilon_{\rm od}=0.7\varepsilon$.
Thick, solid lines are a guide for the eye.
(b) The interfacial energy, $E_{\rm rd}$, between the red substrate and the droplet and the interfacial
energy, $E_{\rm od}$, between the orange substrates and the droplet as a function of time, $t$, for the
systems A and B, as indicated. Also, the instantaneous velocity of the centre of mass of the droplet in
the $x$ direction (cf. Figure~\ref{fig:1}), $\varv$, is plotted for the systems A and B, as indicated.
The dashed, horizontal line corresponds to $\varv=0.0\sigma/\tau$ and the corresponding values of the
velocity are indicated on the right $y$ axis of the graph. Negative values of $\varv$ indicate that
the droplet moves to the left, while positive values indicate that the droplet moves towards the positive
direction of the $x$ axis of our coordinate system. (c) Same as in panel (b), but the systems C and D are
shown, as indicated. Lower panels (d--g) show snapshots at particular times for systems A and D
which are highlighted by a black circle on the graphs of panels (b) and (c) and indicated by the letter of
the corresponding panel (d--g). The interfacial energies, $E_{\rm rd}$ and $E_{od}$, which are relevant
for our discussion in each case are shown below the snapshots for each case. }
\end{figure*}

Figure~\ref{fig:3} illustrates results that indicate whether droplets of different sizes (small, $N=1120$
beads; medium, $N=10080$ beads; large, $N=50400$ beads) can overcome a certain barrier of height $H$.
It suggests that the cases with $\varepsilon_{\rm rd}=\varepsilon_{\rm od}$ will always lead to
pinned droplets irrespective of the droplet size. This is merely due to the physical pinning
barrier, which, albeit small (\textit{e.g.} values as low as $H=\sigma$), is enough to hinder the beads attached to the substrate at the contact line
to climb onto the orange substrate. Moreover, as discussed in the context of
Figure~\ref{fig:2}, $\varepsilon_{\rm od}$ should always be larger than $\varepsilon_{\rm rd}$ for
depinning to take place. In addition, the results of Figure~\ref{fig:3} indicate clearer that a larger
difference between $\varepsilon_{\rm od}$ and $\varepsilon_{\rm rd}$ allows for the translocation of the
droplet at higher values of $H$. A choice of $\varepsilon_{\rm od}$ as high as possible is desirable in
order to favour depinning (also, for intermediate values of $\varepsilon_{\rm rd}$ and
$\varepsilon_{\rm od}$), as suggested by Figure~\ref{fig:3}. Considering the case
($\varepsilon_{\rm rd}=0.3\varepsilon$, $\varepsilon_{\rm od}=0.7\varepsilon$), which enables barrier
crossing for the highest values of $H$ and clearly highlights the different areas in the graphs of
Figure~\ref{fig:4}, we can see that low pinning barriers, $H$ (e.g. $H<8\sigma$), will be overcome by all
droplets, independently of their size. However, when $H>7\sigma$, the medium and large droplets will only
be able to cross the pinning barrier, $H$, while the small droplets will remain pinned. As $H$ further
increases, the medium-size droplets will remain pinned when $H>13$, whereas the large droplets
($N=50400$ beads) will still be able to cross the pinning barrier. Finally, for $H>21$, the large
droplets will also remain pinned being unable to overcome the pinning barrier. Hence, our results
suggest that we can separate droplets of different sizes or control their motion in different directions
by properly choosing the wettability of the red and the orange substrates (maximising
$\varepsilon_{\rm od}$ is desirable) and the height, $H$, of the pinning barrier. This approach could take
place in multiple steps, where the small droplets will remain pinned at small $H$. Then, the medium-size
droplets will remain pinned at higher $H$ values and, finally, the larger values will remain pinned at
higher $H$. Of course, different pinning barriers can be applied in different directions, in this way implementing a binary code, where certain droplets can either cross or not the pinning barrier. Our work clearly shows that the different behaviours are distinct and can be achieved by the different choice of parameters.

In Figure~\ref{fig:4}, we provide details on the translocation mechanism of the droplet upon depinning as
the droplet moves from the red substrate towards the top of the parallel orange substrate. For our
discussion, we have selected four specific systems (see the caption of Figure~\ref{fig:4}), but our
conclusions are valid for all the successful depinning cases of Figures~\ref{fig:2} and
\ref{fig:3}. The observed phenomena are dominated by the interfacial interactions, therefore the
analysis of the different interfacial energy components, as well as the total energy of the system should be investigated. 
In fact, the energy of the system provides the information for its most favourable state
(towards equilibrium) for a particular set of parameters (\textit{e.g.} $H$, $\varepsilon_{\rm rd}$,
$\varepsilon_{\rm od}$, and $N$), since the temperature remains constant throughout the simulation, while
no changes in entropy are expected for the droplet and the substrate during the simulation.
In particular, we show the pair potential interaction energy, $E$,
 for the selected systems, the interfacial energy between the droplet and the red substrate, $E_{\rm rd}$, as
well as the interfacial energy between the droplet and the orange substrates, $E_{\rm od}$ (Figure~\ref{fig:4}a). 
In fact, the latter interfacial contributions play the
most important role in this translocation process. Indeed, these interfacial energies show significant
deviations during the crossing of the pinning barrier (Figures~\ref{fig:4}b and c), which also arises from the wettability difference between the substrate. In particular, the
ability of the droplet to establish more interactions (contacts between beads) with the orange substrates
will eventually determine whether the droplet will be able to fully cross a pinning barrier of height $H$.

A closer look at the interfacial energies, $E_{\rm rd}$ and $E_{\rm od}$, provides more details on the
mechanism of the barrier-crossing process (Figures~\ref{fig:4}b and
\ref{fig:4}c). During the crossing of the pinning barrier by the droplet, we observe that the energy
$E_{\rm rd}$ gradually increases (its absolute value decreases, which means less contacts between the
droplet beads and the beads of the red substrate). In contrast, $E_{\rm od}$ gradually decreases (faster
decrease than the increase in $E_{\rm rd}$, also, due to the fact that
$\varepsilon_{\rm rd} < \varepsilon_{\rm od}$), which manifests as an increasing number of contacts between
the droplet and the orange substrates. At a specific time, for example, the one marked by the letter `d' in
Figure~\ref{fig:4}b for system A and in the snapshot of Figure~\ref{fig:4}d, the two interfacial energies
will be equal. In fact, $E_{\rm rd}$ and $E_{\rm od}$ will be equal for all systems at a certain time while
crossing the pinning barrier. However, this happens very early in the depinning process when the difference
between the parameters $\varepsilon_{\rm rd}$ and $\varepsilon_{\rm od}$ is large, as, for example, in the
case of $\varepsilon_{\rm rd}=0.3\varepsilon$ and $\varepsilon_{\rm od}=0.7\varepsilon$ (systems B, C, D).
We underline that $\varepsilon_{\rm od}$ should always be larger than $\varepsilon_{\rm rd}$ in order for the
droplet to be able to cross the pinning barrier, as seen, for example, from our results in
Figure~\ref{fig:2}. On the contrary, when the wettability difference between the substrates is small
(for example in the case of system A, $\varepsilon_{\rm rd}=0.5\varepsilon$ and
$\varepsilon_{\rm od}=0.7\varepsilon$), $E_{\rm rd} = E_{\rm od}$ at later times and when
the droplet has considerably moved over the pinning barrier. In particular, when the parameters
$\varepsilon_{\rm rd}$ and $\varepsilon_{\rm od}$  differ only by $0.1\varepsilon$, then
$E_{\rm rd}=E_{\rm od}$ takes place when the droplet's centre of mass is half way along the pinning
barrier. Hence, the ability to choose the height of the pinning barrier, $H$ and the wettability of the
red and orange substrates provides further possibilities for controlling the position of the droplet
around the pinning barrier, in the cases that the droplet would remain pinned.

We now turn our attention to the dynamics of the droplet motion during the depinning process.
At the initial stages of the barrier crossing, the instantaneous velocity of the centre of mass of the
droplet in the $x$ direction, $\varv$, increases, as the droplet seeks to establish more favourable contacts
with the orange substrates (Figures~\ref{fig:4}b and c). However, as the droplet moves further along the
pinning barrier, the competition between the red and the orange substrates to establish contacts with the
droplet becomes higher since the droplet needs to climb up the pinning barrier in order to create new
contacts with the top orange substrate. At this stage of the barrier crossing, the droplet moves back and forth and slowly drifts
over the pinning barrier. After this stage and as the droplet moves further over the pinning barrier and
because of the higher attraction of the droplet to the orange substrates ($\varepsilon_{\rm od}$ is always
larger than $\varepsilon_{\rm rd}$), $E_{\rm rd}$ will become zero at some point in time and the droplet
will lose its contact with the red substrate. For example, see point `e' in Figure~\ref{fig:4}c and the
corresponding snapshot in Figure~\ref{fig:4}e for system $D$, which illustrates this effect. At this stage
of the translocation process, the droplet is not anymore dragged by the red substrate and is `free' to
establish further contacts with the top orange substrate. The absence of the attraction between the droplet
and the red substrate leads to the increase of the instantaneous velocity, $\varv$, of the droplet, which,
also, translates into the loss of some contacts with the orange substrate, as the droplet tries to obtain
again its spherical-cap shape. This results in an increase of the energy, $E_{\rm od}$, which is marked in
Figure~\ref{fig:4}c with the letter `f'. A snapshot that corresponds to this situation is presented in
Figure~\ref{fig:4}f. A similar behaviour has been discussed in the context of substrates with
heterogeneity, where hysteresis builds up when the strength of the defect is above a certain threshold,
which depends on the contributions of the elastic energy of the droplet and the barrier energy
\cite{Joanny1984}, which is strictly valid when gravitational effects are negligible \cite{Marsh1993}.
After this point, the droplet has managed to overcome the pinning barrier and climb
on top of the orange substrate. However, the droplet has not yet completely reached its equilibrium shape.
For example, the snapshot in Figure~\ref{fig:4}f clearly manifests this situation, since the advancing and
receding contact angles of the droplet considerably differ. The droplet and generally the system as a whole
will reach its final equilibrium state when it will establish a larger number of contacts with the
parallel to the $x-y$-plane, orange substrate. Then, as also $\varv$ indicates, the droplet will move back
and forth on the top substrate and will not return back to establish contacts with the perpendicular orange
substrate or the red substrate. The number of interfacial contacts between the droplet and the substrate must
always be maximised in order the system to minimise its energy, which occurs only when the droplet eventually `sits' on the top substrate. Hence, the
snapshot of Figure~\ref{fig:4}g (highlighted with the letter `g' in Figure~\ref{fig:4}c) is a typical
equilibrium state of any system that can successfully overcome the pinning barrier. This conclusion is very
important, for example, in a droplet separation process since it guarantees that the droplets that
cross the pinning barrier will not return back to the red substrate. The description of the depinning
mechanism, which we have provided here, is the same for all systems that cross the pinning barrier. However,
for small $H$, the peak `f' of Figure~\ref{fig:4} becomes less pronounced, as can be already hinted by
comparing the results for the systems of Figure~\ref{fig:4}. The same is true when the size of the droplet
becomes smaller. Finally, we have mentioned that maximising the wettability of the orange substrates
(large value of the parameter $\varepsilon_{\rm od}$) is desirable in order to overcome ever higher pinning
barriers (cf.\ Figures~\ref{fig:2} and \ref{fig:3}). We have concluded that the minimisation of the
interfacial energy, $E_{\rm od}$, is the driving force that enables the droplet to cross the pinning barrier.

\begin{figure}[bt!]
\centering
 \includegraphics[width=0.5\textwidth]{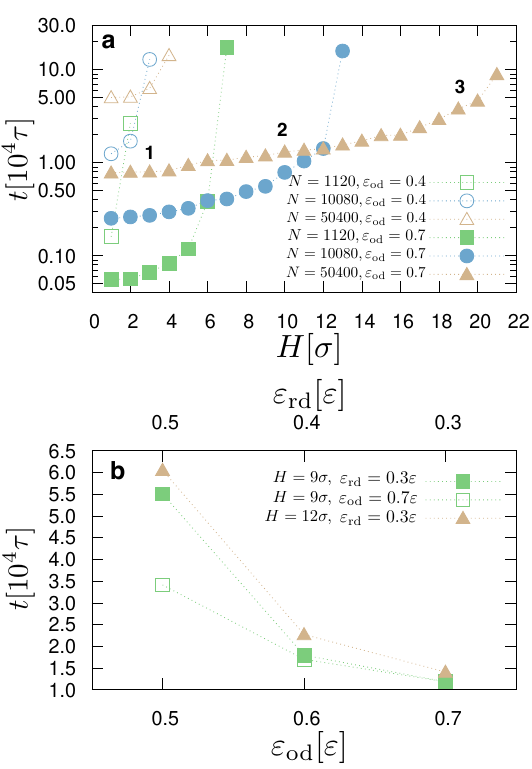}
\caption{\label{fig:5} (a) Time required to cross the pinning barrier as a function of its height, $H$,
for systems with $\varepsilon_{\rm od}=0.4\varepsilon$ (open symbols) or
$\varepsilon_{\rm od}=0.7\varepsilon$
(filled symbols) for different droplet size ($N=1120$: squares; $N=10080$: circles, $N=50400$: triangles),
as indicated. $\varepsilon_{\rm rd}=0.3\varepsilon$ for all cases.
(b) The time required to cross the pinning barrier as a function of $\varepsilon_{\rm od}$ (lower $x$ axis)
when $\varepsilon_{\rm rd}=0.3\varepsilon$ (filled symbols) or as as a function of $\varepsilon_{\rm rd}$
(upper $x$ axis)
when $\varepsilon_{\rm od}=0.7\varepsilon$ (open symbols), as indicated. Cases of different height, $H$,
are shown, as indicated. The choice of these cases is based on the graphs of Figure~\ref{fig:3},
by choosing $H$ values that allow the droplet to cross the pinning barrier for a range of values for
the $\varepsilon_{\rm rd}$ and $\varepsilon_{\rm od}$ parameters.}
\end{figure}

Figure~\ref{fig:5} presents results for the time required by the droplet to cross the pinning barrier. For
the sake of our discussion, we show results of systems with very efficient barrier crossings, that is the
difference in the wettability between the red and the orange substrates is maximised. Hence,
$\varepsilon_{\rm rd}=0.3\varepsilon$ and $\varepsilon_{\rm od}=0.7\varepsilon$. We contrast this behaviour
with the systems that exhibit the least efficient barrier crossings, i.e., systems that can reach
small $H_{\rm max}$ having a small difference in wettabilities, such as the choice
$\varepsilon_{\rm rd}=0.3\varepsilon$ and $\varepsilon_{\rm od}=0.4\varepsilon$. We have also considered
different droplet sizes, as indicated in Figure~\ref{fig:5}a. Overall, all cases show that the time to
cross the pinning barrier increases with the height $H$. While this dependence is monotonic, different
behaviour regimes can be observed. In particular, in the case of $N=50400$,
$\varepsilon_{\rm rd}=0.3\varepsilon$ and $\varepsilon_{\rm od}=0.7\varepsilon$ (Figure~\ref{fig:5}a),
we can clearly discern three regimes. At the first regime (1, Figure~\ref{fig:5}a) for small
values of $H$, namely $\sigma< H < 5$, the effect of the pinning barrier in the translocation process is
very small, due to the large size of the droplet. In this case, the increase of the barrier height, $H$,
does not significantly affect the time that the droplets need to cross the pinning barrier. However,
as the barrier, $H$, further increases, its effect on the time is more tangible, reflecting longer times
that the droplet needs to cross the pinning barrier. This is the second regime (2, Figure~\ref{fig:5}a)
characterised by an exponential growth in time. As we will see by comparison with the different
cases of Figure~\ref{fig:5}, this exponent depends on both the size of the droplet and the particular
choice of the parameters $\varepsilon_{\rm rd }$ and $\varepsilon_{\rm od }$. Hence, it is not possible to
find a universal exponent for the crossing time, but we can observe that this exponent becomes smaller as
the size of the droplet increases. The regime (1) may be a limiting case of this exponent when the effect
the pinning barrier becomes negligible on the time for the droplet to cross the barrier. In the third regime
(3, Figure~\ref{fig:5}a), the droplet takes even more time to cross the pinning barrier and as $H$
increases this time practically becomes infinite. This behaviour reflects the great difficulty of the
droplet to further establish energetically favourable contacts with the top orange substrate. The above
picture for the largest droplet ($N=50400$ beads) also seems to apply in the case of smaller
droplets (\textit{i.e.} $N=1120$ and $N=50400$), of course when the crossing is possible, but with the exception that the behaviour of regime (1)
is absent. This simply means that values as low as $H=\sigma$ already have an important influence on the
translocation process in the case of the small droplets. This impact becomes even higher when the
wettability difference between the red and the orange substrates is small (for example,
$\varepsilon_{\rm rd}=0.3\varepsilon$ and $\varepsilon_{\rm od}=0.4\varepsilon$ as shown in
Figure~\ref{fig:5}a). In this case, some of the droplets are already exhibiting the behaviour of regime
(3), and the times to cross the pinning barrier increase by almost an order of magnitude for certain $H$.
Hence, we conclude that larger droplets offer better control in the time-scale of the process, when
this is relevant for the application design. Our analysis is of course relevant in the absence of 
gravitational effects, that is length scales smaller than the capillary length, which is indeed the case
in our \textit{in silico} experiments.

Finally, we discuss how the time scale of the barrier crossing is affected by changes in the wettability
between the substrates. Based on the results of Figures~\ref{fig:2} and \ref{fig:3}, we consider the cases
shown in Figure~\ref{fig:5}b, for which we can observe barrier crossing for a wide range of parameters
$\varepsilon_{\rm rd}$ and $\varepsilon_{\rm od}$ for fixed $H$. From the results of
Figure~\ref{fig:5}b, we can conclude: Firstly, choosing higher $\varepsilon_{\rm od}$ values leads to
faster barrier crossings for the same difference between substrates wettability as expressed through the
parameters $\varepsilon_{\rm rd}$ and $\varepsilon_{\rm od}$. Secondly, higher $H$ values appear to affect
proportionally the time of crossing the pinning barrier across the range of parameters
$\varepsilon_{\rm rd}$ and $\varepsilon_{\rm od}$. Our conclusions seem to apply throughout the
systems of this study. However, a more comprehensive discussion would still require larger droplets
than the ones considered here, which goes beyond our current computational capabilities and the scope of
this work.

\section{CONCLUSION}
In this study, we have investigated the pinning of liquid droplets on solid substrates. We have discussed the necessary
conditions for pinning and the mechanism of crossing the pinning barrier upon depinning.
We found that even the smallest barrier, namely $H=\sigma$, is able to keep the droplet pinned
when the wettability of the physical barrier is equal or smaller than the wettability of the substrate where
the droplet `sits' before crossing the barrier. This is true for all droplet sizes considered in our study. Moreover, the
crossing of a higher pinning barrier ($H_{\rm max}$) by the droplet is favoured by a larger wettability
of the substrates that form the barrier (orange substrates). In such cases, the crossing of
the barrier will also be quicker. The time
scale of the crossing depends on the size of the droplet, $N$, and the wettability of the substrates as
expressed through $\varepsilon_{\rm rd}$ and $\varepsilon_{\rm od}$. In addition, we found that larger
droplets can cross higher pinning barriers. We have analysed in detail the mechanism of the barrier
crossing and have identified the driving force of this process, which is the minimisation of the system's
energy, with the main contribution coming from the decrease of the interfacial energy, $E_{\rm od}$,
between the orange substrate and the droplet. To this end, we have presented a detailed discussion of the
pinning--depinning mechanism and the barrier crossing by the droplet, and we have analysed the dynamics
of this process based on the instantaneous velocity of the centre of mass of the droplet and the time-scale
of the crossing. For dynamics of movement, we have identified three different time-scale regimes and discussed its
implications for applications exploitation. Furthermore, we have also described how pinning and depinning processes can be exploited in nanotechnology
applications by controlling the droplet motion through a proper choice of the pinning barrier and the
substrate wettabilities of the red and orange substrates for a given droplet size. Our study provides
ways of separating and steering droplets on solid substrates. In this way, we anticipate that our work
could have direct implications in various nanotechology applications, especially in the areas of
microfluidics, microfabrication, coatings, and biology.

\begin{acknowledgement}

This project has received funding from the European Union's Horizon 2020 research and innovation programme
under the Marie Sk{\l}odowska-Curie grant agreement No.\ 778104. This research was supported in part by PLGrid
Infrastructure.

\end{acknowledgement}

\begin{suppinfo}

Movies of barrier crossing by the droplet: 
M1*: $H=3.0\sigma$, $\varepsilon_{\rm rd}=0.3\varepsilon$, $\varepsilon_{\rm od}=0.5\varepsilon$, $H=1120$ beads;
M2*: $H=10.0\sigma$, $\varepsilon_{\rm rd}=0.3\varepsilon$, $\varepsilon_{\rm od}=0.7\varepsilon$, $H=10080$ beads;
M3*: $H=15.0\sigma$, $\varepsilon_{\rm rd}=0.3\varepsilon$, $\varepsilon_{\rm od}=0.6\varepsilon$, $H=50400$ beads;
A movie of a pinned droplet (no barrier crossing) for the sake of comparison:
M4*: $H=11.0\sigma$, $\varepsilon_{\rm rd}=0.3\varepsilon$, $\varepsilon_{\rm od}=0.4\varepsilon$, $H=50400$ beads;

\end{suppinfo}

\bibliography{bib_pinning}

\end{document}